\documentclass[conference]{IEEEtran}
\IEEEoverridecommandlockouts

\usepackage{cite}
\usepackage{amsmath,amssymb,amsfonts}
\usepackage{cite}
\usepackage{graphicx}
\usepackage{textcomp}
\usepackage{xcolor}
\usepackage[acronym]{glossaries}
\usepackage{mathtools}
\usepackage{makecell}
\usepackage{multirow}
\usepackage{tikz}
\usepackage{siunitx}
\def\BibTeX{{\rm B\kern-.05em{\sc i\kern-.025em b}\kern-.08em
    T\kern-.1667em\lower.7ex\hbox{E}\kern-.125emX}}

\newcommand*\circled[1]{\tikz[baseline=(char.base)]{
            \node[circle,draw,inner sep=2pt, thick, text=black] (char) {#1};}}

\newacronym{xr}{XR}{Extended Reality}
\newacronym{slam}{SLAM}{Simultaneous Localisation and Mapping}
\newacronym{ate}{ATE}{Absolute Trajectory Error}
\newacronym{imu}{IMU}{Inertial Measurement Unit}
\newacronym{rgbd}{RGB-D}{RGB-Depth}
\newacronym{ber}{BER}{Bit Error Rate}
\newacronym{snr}{SNR}{Signal-to-Noise Ratio}

\renewcommand{\baselinestretch}{.98}
\IEEEsettopmargin{t}{0.8in}

\begin{document}

\title{
Multi-User XR Offloading via Massive MIMO: A System-Level Analysis using a Real-Life Dataset
\thanks{This research has been supported by the Swedish Foundation for Strategic Research (CSS22-0003) through the Chalmers-Lund Center for Advanced Semiconductor System Design (classIC).}
}

\author{
Love Bárány, Ilayda Yaman, Ove Edfors, Amir Aminifar, Liang Liu\\
\textit{Dept. of Electrical and Information Technology, Lund University, Sweden}\\
Email: \texttt{first.last@eit.lth.se}
}

\maketitle

\begin{abstract}
\gls{slam} is one of the biggest bottlenecks of \gls{xr} devices, which have strict requirements for latency, power consumption, and user satisfaction.
A solution that has been proposed and studied to meet the requirements is to offload \gls{slam} to a remote server, which leverages computational hardware but may suffer due to incurred delays and transmission power.
In this work, we propose offloading \gls{slam} using Massive MIMO, which is attractive due to lower latencies, transmission power, and a more reliable link for multiple users. 
A framework for system-level analysis of latency and localisation error in multi-user offloaded \gls{xr} with Massive MIMO has been proposed, and a case study with varying system-level parameters has been performed with it.
The case study showed that there are important trade-offs between latency, localisation error, and device transmission power.
We find that Massive MIMO is a promising technology for \gls{xr} offloading, but that further evaluations including complete device power consumption are needed to get the full picture.
\end{abstract}

\glsresetall
\section{Introduction}

\gls{xr} devices contain computationally demanding blocks, but have strict constraints on latency, power consumption, and user satisfaction.
One of the most demanding blocks in \gls{xr} devices is vision-based \gls{slam}, which can occupy more than half the available CPU time \cite{huzaifaILLIXROpenTestbed2022}. 
\gls{slam} algorithms use sensor inputs, such as camera images, to jointly estimate the position of the sensor and a map of the world around it.

A solution to meet the latency constraints, while keeping power consumption down, is to offload demanding computations to a remote server with computational abilities.
In previous offloaded \gls{slam} evaluations, the offloaded portion has been placed on an edge or cloud server, and offloading is done through Wi-Fi or 5G links \cite{jiangRemoteVIOOffloadingHead2025, benaliEdgeSLAMEdgeAssistedVisual2022,  
zhangOrchestratingJointOffloading2025,sossallaDynNetSLAMDynamicVisual2022, panRobustCollaborativeVisualInertial2024a}. Some evaluate multi-user scenarios \cite{panRobustCollaborativeVisualInertial2024a}, while most evaluate a single-user scenario \cite{jiangRemoteVIOOffloadingHead2025, benaliEdgeSLAMEdgeAssistedVisual2022, sossallaDynNetSLAMDynamicVisual2022, zhangOrchestratingJointOffloading2025}. Furthermore, most only evaluate one partitioning of work between the device and the remote server \cite{jiangRemoteVIOOffloadingHead2025, benaliEdgeSLAMEdgeAssistedVisual2022, panRobustCollaborativeVisualInertial2024a, zhangOrchestratingJointOffloading2025}.

Although offloading can reduce latency and power by leveraging powerful hardware, it introduces further latency and power consumption due to wireless transmissions.
If the link quality is poor and data is lost or corrupted, user satisfaction may be negatively affected.
Additionally, traditional communication schemes may not allow all users to transmit at all times using the full bandwidth, meaning users either have to wait or only use a subset of available frequencies.

Massive MIMO is a mobile communication paradigm in which the number of antennas at the base station is much larger than the number of users that the antennas serve.
Users can transmit data simultaneously at all times, using the full bandwidth, via spatial multiplexing.
The Massive MIMO communication link is also more reliable due to channel hardening \cite{marzetta_noncooperative_2010}. Furthermore, increasing the number of antennas increases the antenna gain, which can reduce the required transmission power on devices \cite{ngo_energy_2013}.

Thereby, we propose to leverage Massive MIMO for multi-user \gls{xr} offloading, as well as to utilise hardware at the base station for the offloaded computations, which would avoid further routing.
To assess the fit of Massive MIMO with \gls{xr} offloading, it must be determined whether latency constraints are met, whether the link quality is good enough for offloading, and what the impact on device power consumption is.
For power, we will only consider the transmission power required on \gls{xr} devices.
In order to accurately evaluate offloading with Massive MIMO, we use a real-life dataset that contains synchronised images, wireless measurements, and ground truth trajectories to evaluate \gls{slam} accuracy: the Lund University Vision, Radio, and Audio (LuViRA) dataset \cite{yamanLuViRADatasetValidation2024}.

Our main contribution in this paper is a framework for system-level analysis of multi-user offloaded \gls{xr} with Massive MIMO.
This framework is then used in an exploratory study to evaluate the trade-offs between latency, localisation error, and device transmission power for three offloaded scenarios with varying wireless frame structures and \glspl{ber}.

\begin{figure*}
    \centering
    \includegraphics[width=\linewidth]{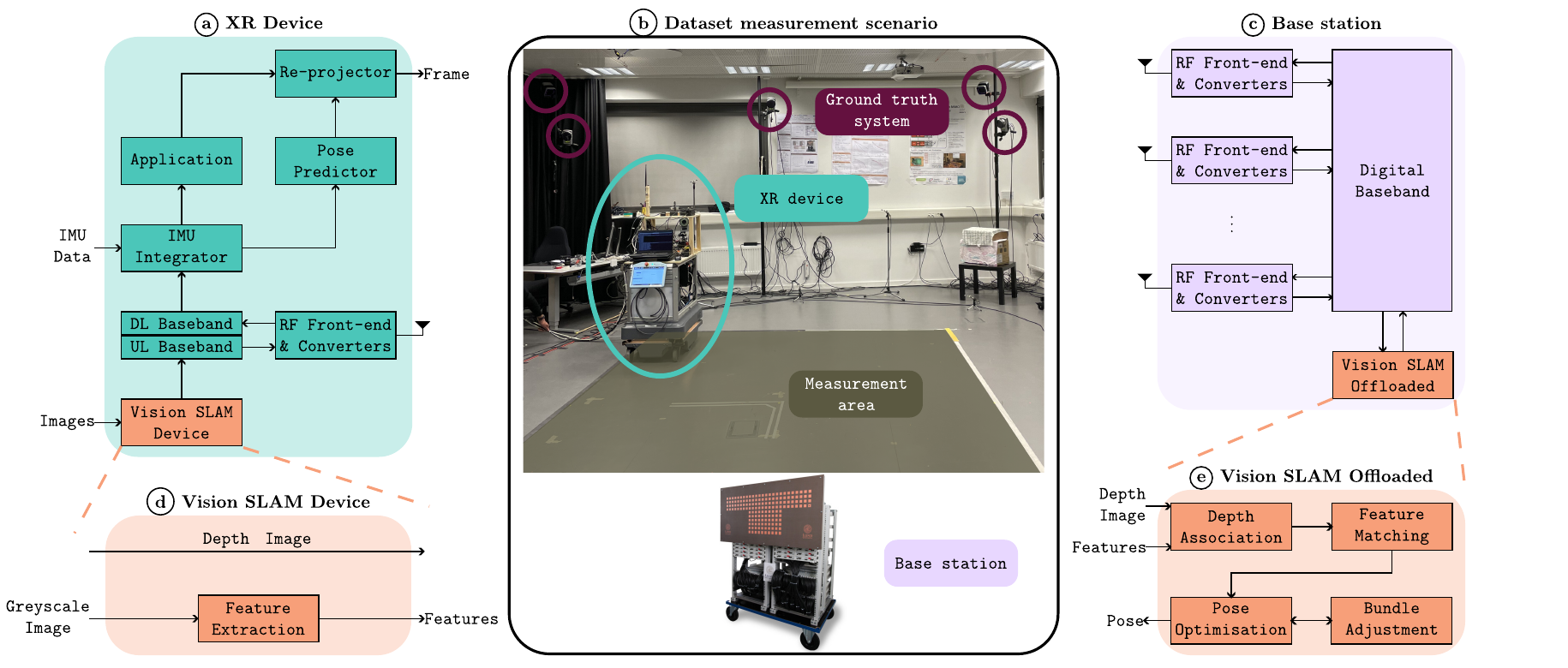}
    \caption{Overall offloaded \gls{xr} diagram. Vision SLAM components are simplified for illustrative purposes, and are only an example of the specific tasks and partitioning between device and offloaded.}
    \label{fig:sys-model-diag}
\end{figure*}

\section{Background and system overview}

In \gls{xr} devices, an \gls{imu} and pose predictor provide fast pose estimates to meet display latency requirements. \gls{imu} estimates are known to drift over time, which \gls{slam} corrects. If the \gls{slam} correction takes too long, drift will build up, and user satisfaction will be negatively affected due to inaccurate pose estimates.
User studies indicate that user satisfaction is maintained if the latency of pose correction is below $200\unit{\milli\second}$, with some allowed variability \cite{jiangRemoteVIOOffloadingHead2025}. 

Figure~\ref{fig:sys-model-diag} shows the overall function blocks for offloaded vision \gls{slam} using massive MIMO, with the \gls{xr} device model \textbf{\circled{a}}, the LuViRA dataset measurement setup \textbf{\circled{b}}, and the Massive MIMO base station model \textbf{\circled{c}}.

Measurements in the LuViRA dataset were taken in an area of size $4.2 \times 2.5$m$^2$. The base station used is the Lund University Massive MIMO (LuMaMI) testbed \cite{malkowskyWorldsFirstRealTime2017}, which implements fully digital beamforming with 100 antennas and 1200 subcarriers. Transmissions occur at a central frequency of 3.7GHz, with a bandwidth of 20MHz.
The user equipment is equipped with an Intel RealSense D435i depth camera providing colour and depth images, an \gls{imu} for inertial sensor data, and an antenna for wireless channel state information.
The ground-truth system uses motion capture cameras to measure the true trajectory of the user equipment, with a sub-millimetre accuracy \cite{yamanLuViRADatasetValidation2024}.

In order to simulate a multi-user scenario, we concatenate massive MIMO channel matrices of 10 trajectories to mimic a scenario where 10 users are offloading simultaneously, all using the full 20MHz bandwidth. This concatenated matrix is then used to simulate \gls{ber} for all 10 users, in terms of the post-equalisation \gls{snr} received at the base station per user. The multi-user aspect of this work is captured through this realistic channel model.
It is assumed that all \gls{xr} devices transmit with power control---the devices dynamically adjust their transmission power so that the \gls{snr} received at the base station is the same for all users, regardless of their distance from it.

As shown in Figure~\ref{fig:sys-model-diag}, vision \gls{slam} is divided into vision \gls{slam} device \textbf{\circled{d}} and vision \gls{slam} offloaded \textbf{\circled{e}}.
The exact task division between the device and the base station differs depending on the offloading scenario.
The tasks in the vision \gls{slam} blocks are based on the \gls{rgbd} pipeline of ORB-SLAM3\cite{camposORBSLAM3AccurateOpenSource2021}.
First, features are extracted from the greyscale image. The features are then assigned a relative depth from the camera, using the depth image, to create 3D coordinates. The extracted features are matched with features found in previous images, and the matches are used to estimate the pose in pose optimisation. The estimated pose and features are then further optimised with bundle adjustment, to improve the consistency of the estimated trajectory and enabling more accurate future estimations.

In our exploratory study, we will assume that both \gls{slam} device and offloaded are computed in the physical layer, without packeting beyond mapping to OFDM symbols and without error correction. While error correction ensures that user satisfaction does not suffer, it introduces latency to detect and correct the errors, with a similar trade-off for more packeting. A real system would most likely contain some form of the two; however, we wish to evaluate how sensitive the \gls{slam} accuracy is to raw bit errors. The results of this sensitivity study can then be used to guide the amount of error correction needed.

\section{Analysis framework}

To evaluate the system, we developed an analysis framework for evaluating pose correction latency and the sensitivity of \gls{slam} accuracy to bit errors in terms of localisation error. The framework focuses on \gls{slam} execution and massive MIMO communication during offloading. It combines real-world implementation and measurement results with analytical models, making the framework both realistic and flexible for exploring design trade-offs.

\subsection{Latency model}\label{sec:lat-model}

We define the pose correction latency as the time between images being input to a pose being computed and transmitted back to the device, and model it as:
\begin{equation}\label{eq:lat}
    \tau_\text{pose} = \tau_\text{device}+\tau_\text{UL}+\tau_\text{BS}+\tau_\text{offloaded}+\tau_\text{DL}.
\end{equation}
$\tau_\text{device}$ and $\tau_\text{offloaded}$ are, respectively, the execution times of the \gls{slam} algorithm on the device and the offloading base station. Their sum is the total latency of the \gls{slam} task.
$\tau_\text{UL}$ and $\tau_\text{DL}$ are the latencies for transmitting the offloaded data to and from the base station in terms of the radio frame structure, respectively.
$\tau_\text{BS}$ is the processing latency of the RF front-ends and digital baseband on the base station.

The latency model is a hybrid model in that $\tau_\text{device}$, $\tau_\text{offloaded}$, and $\tau_\text{BS}$ come from experimental results, while $\tau_\text{UL}$ and $\tau_\text{DL}$ are model-based.
For $\tau_\text{BS}$, we use an FPGA implementation result from the LuMaMI testbed that $\tau_\text{BS}=132\unit{\micro\second}$\cite{malkowskyImplementationLowLatencySignal2016}. 
For $\tau_\text{device}$ and $\tau_\text{offloaded}$, we use experimental results from executing ORB-SLAM3 on an embedded device and a high-end desktop computer, respectively.
A modified ORB-SLAM3 package is used, which records the time when images are input, when offloading occurs, and when a pose is returned.

We express the transmission times as:
\begin{equation}\label{eq:tx}
\begin{split}
    \tau_\text{UL} = &\tau_\text{symb}\cdot(N_\text{symb wait UL} + N_\text{symb/pose UL}+\\
                        & N_{\text{slot/pose UL}}\cdot(N_\text{symb}-N_\text{UL symb})),
\end{split}
\end{equation}
for uplink, replacing UL with DL where appropriate for downlink. A radio frame is organised into a number of slots, with each slot containing $N_\text{symb}$ OFDM symbols. Of these, $N_\text{UL symb}$ are uplink data symbols and $N_\text{DL symb}$ are downlink data symbols. 
Each OFDM symbol is $\tau_\text{symb}=71.4\unit{\micro\second}$ long in the LuMaMi testbed. 
$N_\text{symb/pose [UL;DL]}$ and $N_\text{slot/pose [UL;DL]}$ are, respectively, the number of OFDM symbols and full slots needed to transmit the offloaded data.
$N_\text{symb wait [UL;DL]}$ is the number of symbols to wait before transmission can start in the worst case, defined as in \cite{tinnerberg2025spectrumefficiencyprocessinglatency}.

\begin{figure}
    \centering
    \includegraphics[width=\linewidth]{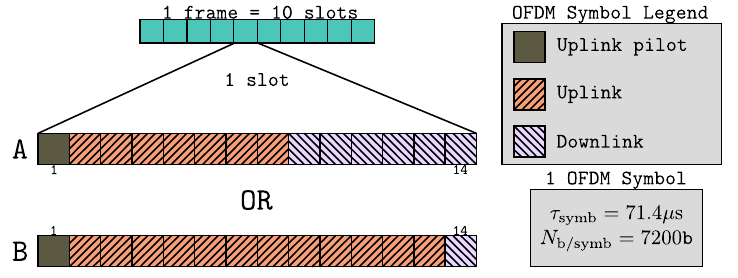}
    \caption{Frame structures compared: A (top) and B (bottom). 1200 subcarriers with 64-QAM modulation is assumed. Pilot symbols are used to estimate the channel at the base station.}
    \label{fig:frame_struct}
\end{figure}

In the offloading scenarios we study, there is an imbalance in the amount of data transmitted between uplink and downlink: uplink consists of sensor inputs at various stages of processing, while downlink consists of computed poses. As such, we would like to investigate the impact of different radio frame structures on the pose correction latency. As a case study, we select the two structures shown in Figure~\ref{fig:frame_struct}. Frame structure A balances uplink and downlink traffic, while frame structure B is optimised for an uplink-heavy scenario.
All users transmit using all OFDM data symbols and subcarriers simultaneously.

Since we perform an exploratory study, we make a few assumptions for simplicity.
We assume that the latencies of the RF front-end, converter, and baseband processing on the device are negligible compared to other latencies in the model. 
Since the scenario considered is indoors, signal propagation latency is also assumed to be negligible.
Finally, we assume that there are enough resources to serve all communication with all users without queueing or other incurred delays.

\subsection{Localisation error model}

In order to quantitatively evaluate the resulting trajectories generated by the \gls{slam} block, \gls{ate} as described in \cite{zhang_tutorial_2018} can be used. 
In this work, only the translation component of \gls{ate} is evaluated, which we refer to as the localisation error.
We use \texttt{evo} \cite{grupp2017evo} to align the estimated trajectories with the corresponding ground truths using all poses in the estimated trajectories, to re-scale, and to compute the localisation error.
As shown in \cite{yamanLuViRADatasetValidation2024}, the baseline localisation error of the trajectories varies. If errors are averaged and compared directly, the average may be biased towards trajectories with larger baseline errors. Furthermore, the change in error with offloading is of greater relevance than the errors themselves.
Therefore, results are normalised by computing the percentage error difference relative to a no-offloading baseline. Errors are first normalised and averaged per trajectory, then averaged over all trajectories.

To quantify the sensitivity of \gls{slam} accuracy to bit errors, we simulate bit errors that would occur during wireless transmission, without error correction. Due to the large number of antennas in Massive MIMO, we assume that bit errors are uniformly distributed rather than appearing in bursts. 
For data that would be offloaded, the number of bit errors that occur is sampled from a binomial distribution with an uncoded \gls{ber} as probability of success. Bit error locations are then sampled uniformly without replacement. Offloaded data that becomes invalid after bit errors is restricted to known allowed ranges, so that the \gls{slam} algorithm does not crash.

\section{Results and discussion}

We will now use the framework to explore the trade-offs between latency, localisation error, and device transmission power, for 3 offloading scenarios, 2 wireless frame structures, and 4 \glspl{ber}. The embedded device is a Jetson Orin Nano 8GB with 2 active CPU cores, configured for minimal power consumption. 
The desktop computer contains an Intel(R) Core i7-2600K CPU and a discrete GeForce GTX 1080 Ti GPU.  

\subsection{Offloading scenarios compared}

We wish to study the effect of transmitting inputs at various stages of processing. 
A more processed input will take less time to be transmitted because it is smaller, but may be more sensitive to bit errors.
The data transmitted and the packet size for the three scenarios considered are given in Table~\ref{tab:bits_per_s}. We can see that there is an imbalance between uplink and downlink traffic, which motivates the comparison of frame structures.
The scenarios differ in task division between the device and the base station. In all scenarios, pose optimisation and bundle adjustment are performed on the base station.
In the first scenario, the device does minimal processing.
In the second scenario, the device extracts features from the greyscale image.
In the third scenario, the device extracts features and assigns a depth value to each.
The partitioning in Figure~\ref{fig:sys-model-diag} corresponds to the second scenario.

\begin{table}
    \centering
    \caption{Summary of offloading scenarios, with type of data transmitted and packet sizes, for uplink and downlink.}
    \label{tab:bits_per_s}
    \begin{tabular}{|c|c|c|c|c|}
        \cline{2-5}
                         \multicolumn{1}{c|}{} & \multicolumn{2}{c|}{Uplink} & \multicolumn{2}{c|}{Downlink} \\\hline
        \textbf{Scenario} & \textbf{Transmitted} & \textbf{\makecell{Packet\\ size (KiB)}} & \textbf{Transmitted} & \textbf{\makecell{Packet\\ size (KiB)}} \\\hline
        1 & \makecell{Greyscale \&\\depth images} & 900 & Pose & 0.04 \\\hline
        2 & \makecell{Features \&\\ depth image} & 672 & Pose & 0.04 \\\hline
        3 & \makecell{Features\\with depths} & 84 & Pose & 0.04 \\\hline
    \end{tabular}
\end{table}

\subsection{Pose correction latency results}

\begin{figure}
    \centering
    \includegraphics[width=\linewidth]{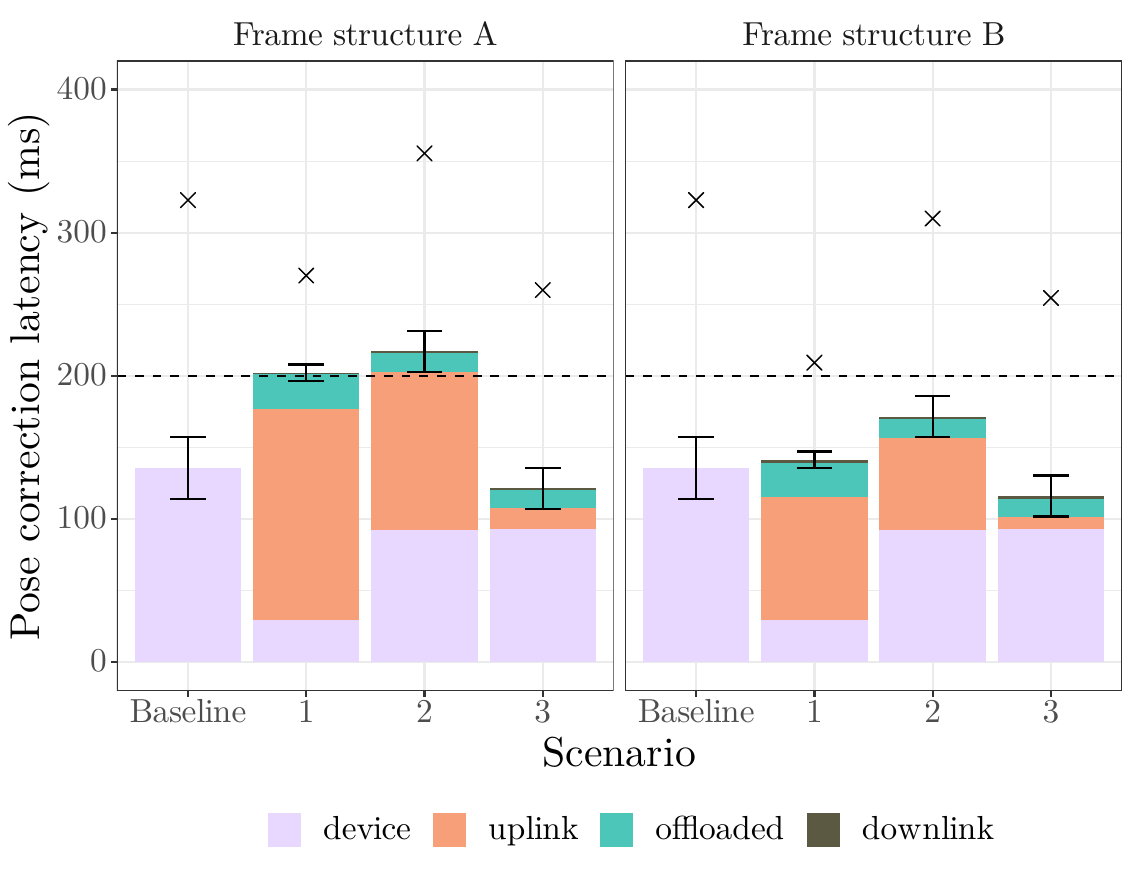}
    \caption{Average pose correction latency by term in \eqref{eq:lat} across all trajectories, by scenario and frame structure. $\tau_\text{BS}$ not visualized due to being small compared to other terms. Error bars indicate standard deviation, while crosses represent the worst case latency.}
    \label{fig:latency}
\end{figure}

Figure~\ref{fig:latency} shows pose correction latency by scenario and frame structure.
We can see that the pose correction deadline of $200\unit{\milli\second}$ is met on average in almost all combinations, but never in the worst cases.
The dominant terms in \eqref{eq:lat} seem to be $\tau_\text{device}$ and $\tau_\text{UL}$, with some variation depending on the scenario. An important observation is that $\tau_\text{device}$ more than doubles when features are extracted on the device in scenarios 2 and 3, compared to scenario 1. As such, we could reduce the overall latency for these scenarios by accelerating feature extraction on the device.
As an example, an accelerator implementing feature extraction could extract 1000 features in $7.1\unit{\milli\second}$\cite{ferreiraEnergyEfficientApplicationSpecificInstructionSet2021}.
Although using frame structure B reduces $\tau_\text{UL}$ somewhat, it may not be the best solution for this. Since it is unlikely that the only transmissions that occur are related to \gls{slam}, other applications with more downlink traffic may be throttled if frame structure B is used. Instead, some type of compression or fitting more bits into each data symbol may be a better solution.

\subsection{Localisation error results}

\begin{figure}
    \centering
    \includegraphics[width=\linewidth]{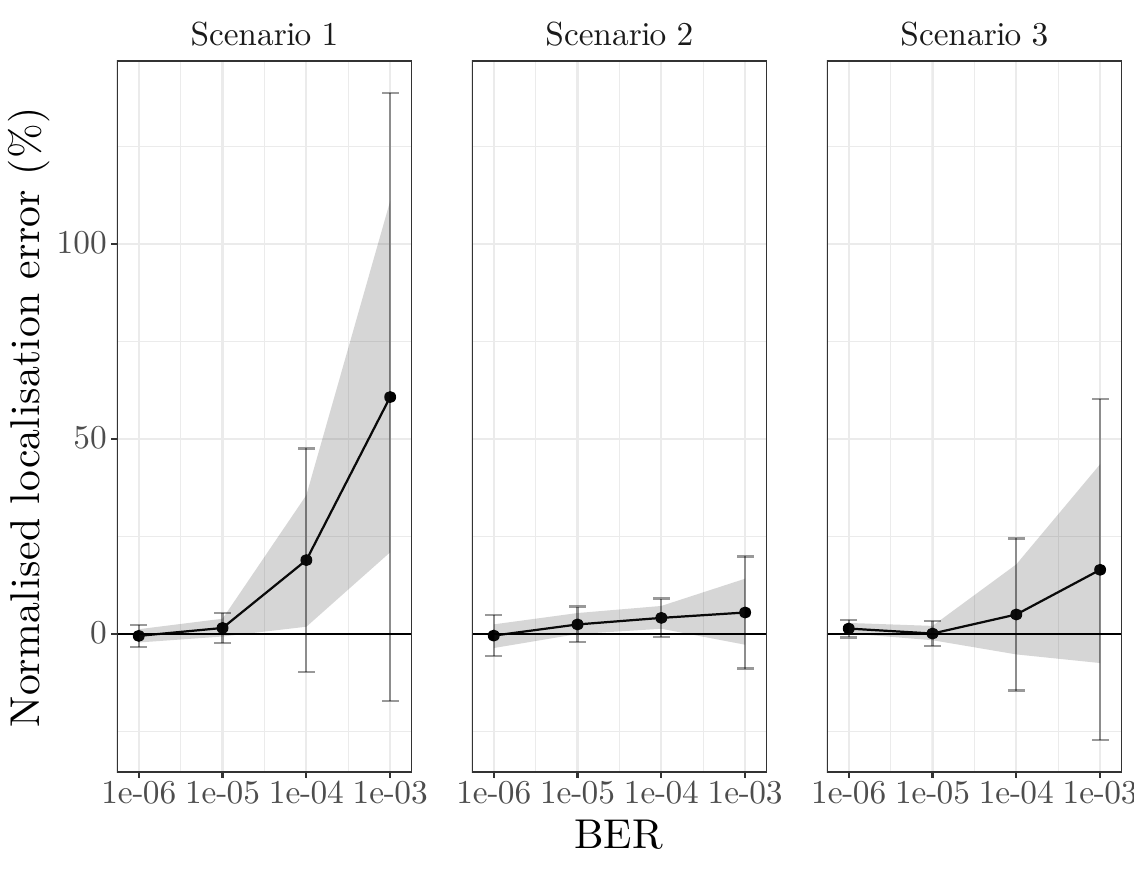}
    \caption{Normalised localisation error versus \gls{ber} by scenario. Bootstrap mean (points) with $95\%$ confidence interval (ribbon) and bootstrap standard deviation (error bars), 10000 draws.}
    \label{fig:per_rmse}
\end{figure}

Localisation error results are presented in Figure~\ref{fig:per_rmse}.
Even if all scenarios follow a trend of increased error with increased \gls{ber}, there is a difference in magnitude and spread. For scenario 2, a \gls{ber} of $10^{-4}$ exhibits lower normalized errors than the other two scenarios, that need a \gls{ber} of $10^{-5}$ to match it. Scenario 2 has the lowest error and the smallest spread for most \glspl{ber}, indicating that it is the most robust scenario of the three in the presence of random bit flips.
This should be studied further, as it is not obvious why features and the depth image are less affected by bit errors compared to the greyscale image and processed depths.
To decrease the \gls{ber}, we could increase the \gls{snr}, which we discuss in the next section. Another possibility is to reduce the number of bits per symbol, but this would lead to an increase in latency.

\subsection{Impact on device transmission power}

\begin{figure}
    \centering
    \includegraphics[width=.8\linewidth]{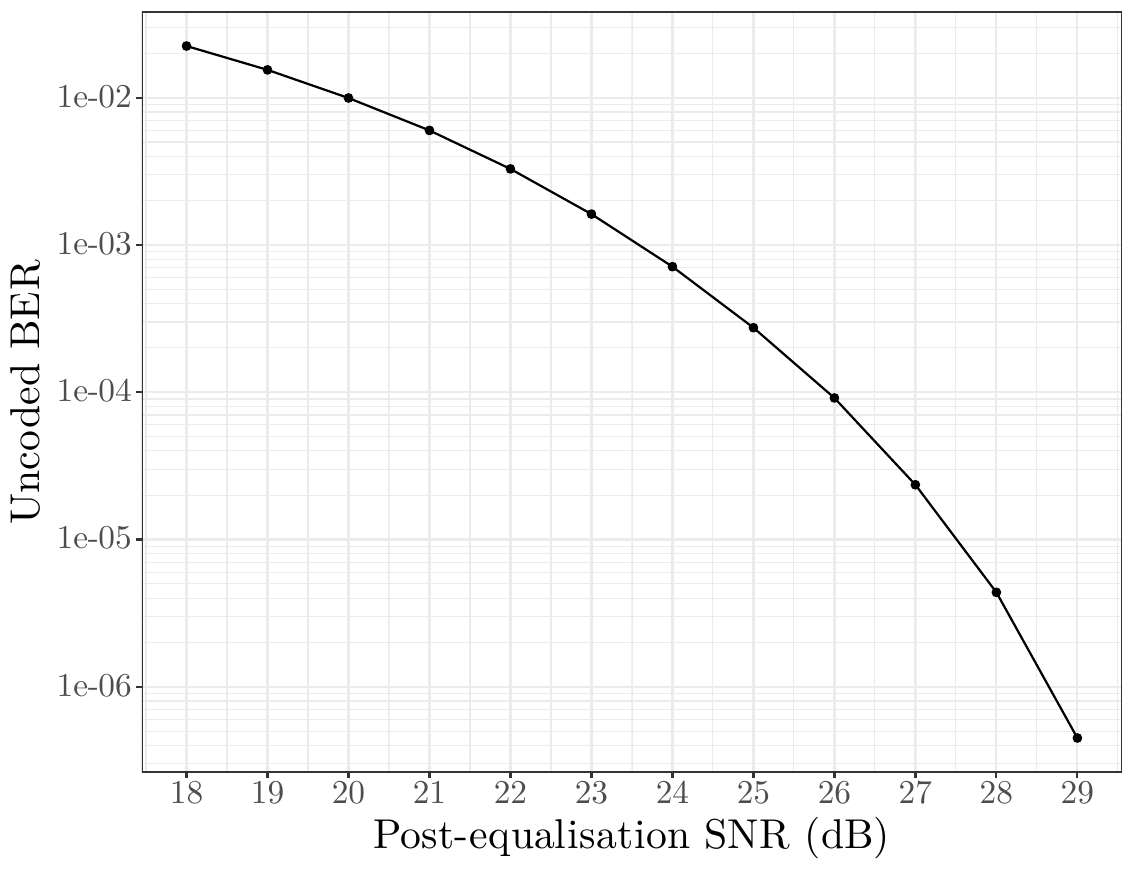}
    \caption{Average uncoded \gls{ber} of Massive MIMO channel with 10 users versus post-equalisation \gls{snr}. Received signals are equalised with zero-forcing.}
    \label{fig:ber-snr}
\end{figure}

From these results, we wish to estimate the required device transmission power. Initially, we perform system \gls{ber} simulations to find the required post-equalisation \gls{snr} per user for a given \gls{ber}, shown in Figure~\ref{fig:ber-snr}.
We consider an indoor factory scenario, with a 
distance between a device and the base station of $100\unit{\meter}$.
We assume thermal noise at $300~\unit{\kelvin}$, a fading margin of $2.5\unit{\decibel}$, and a receiver noise figure of $8\unit{\decibel}$.
For simplicity, we assume free-space path loss, but a more careful analysis for more specific environments can be made.
The central frequency and bandwidth come from the LuMaMI testbed. Using a first-order link budget, we estimate that the transmission power required for an uncoded \gls{ber} of $10^{-4}$ and $10^{-5}$ is $0.856\unit{\milli\watt}$ and $1.356\unit{\milli\watt}$, respectively.
These powers are low, not only because of the number of antennas in Massive MIMO, but also because we only consider very short distances.
With these first-order estimates, we can see a significant relative difference. However, the actual difference is small and may not be significant compared to the power consumed by other components.
Note that these are powers required at the output of a power amplifier; the actual power consumption will depend on the efficiency of the amplifier. 

\subsection{Trade-offs}

There are quite a few trade-offs that become apparent when considering the combination of all results.
If our goal is to decrease the device processing, and thus the device latency, the system becomes more sensitive to bit errors, and the transmission time increases.
In order not to increase the error significantly, a lower \gls{ber} is needed, which requires a higher \gls{snr}. A higher \gls{snr} requires more transmission power.
If we reduce the transmission time by compressing the data, latency for compression will be introduced. Feature extraction and depth association can be considered forms of compression, and as previously seen, the latency more than doubles when performing this compression compared to transmitting raw images. If we instead increase the amount of data per data symbol, intersymbol interference will increase, leading to higher \gls{snr} requirements to achieve the same \gls{ber}. 

If our goal instead is to minimise errors introduced by offloading, we need to have more transmission power, or do more processing on the device. The overall latency is similar or lower than a no-offloading baseline in these cases, but the device processing latency is not much lower than in the baseline. This may imply that power consumption is not reduced. However, we should note that bundle adjustment, which is the most computationally intensive process in ORB-SLAM3, is moved to the base station in all scenarios, and does not have a direct impact on pose correction latency.
Therefore, any conclusion regarding device power consumption cannot be made without a more thorough evaluation. 

With our assumptions, scenario 2 provides the best latency-robustness trade-offs, while scenario 1 minimizes device latency, and scenario 3 minimizes network traffic.

\section{Conclusions}

We have proposed a framework for exploring latency, localisation error, and device transmission power trade-offs in multi-user offloaded \gls{xr} with Massive MIMO.
We set up three task division scenarios and identified the trade-offs for different configurations of scenario, frame structure, and \gls{ber}.
We also compared the offloaded scenarios with a baseline without offloading.
Our work has shown Massive MIMO to be promising as an enabler of offloaded multi-user \gls{xr}, with further studies needed. In this work, the multi-user aspect was captured through the realistic channel, but other effects such as scheduling should be evaluated.
Future work will focus on a more detailed device power consumption model, as well as base station hardware architectures to support both mobile communication tasks and offloaded \gls{xr} computations.

\bibliographystyle{IEEEtran}
\bibliography{IEEEabrv,main.bib}

\end{document}